# A data transmission system for the phase contrast X-ray human computed tomography prototype

Rongqi Sun, Lian Chen, Houbing Lu, Feng Li, and Ge Jin

*Abstract–The cross section of X-ray phase contrast caused by these low-Z elements is greatly bigger than the absorption. Therefore, in the field of X-ray imaging, the phase shift information can offer better imaging contrast. In this paper, we present a data transmission system for the detectors of a phase contrast X-ray human computed tomography prototype. This system contains 3 data collecting boards (DCB) and one data transmitting board (DTB). A slip ring is used to transmit the data from the rotator side to the stator side over a nowadays commonly used multi-mode fiber (MMF). On the rotator side, 3 DCBs act as the controller of these detectors. The function of the DTB is to store all image data from 3 DCBs and implement the commutation with PC. The test shows that this system can meet the requirement of the prototype.*

## I. Introduction

X-ray radiographic imaging relying on absorption contrast is wildly applied into the research areas such as materials science, clinical medicine. However, the absorption coefficient of light elements for X-ray is much lower than heavy elements. For those samples composited of light elements like human soft tissues, the X-ray luminance and imaging exposure time should be stepped up to obtain sufficiently high image quality. This will lead to the sample harmed by the nuclear radiation. Theoretically, the cross section of X-ray phase contrast caused by these low-Z elements is greatly bigger than the absorption. Besides, the acquisition of phase information can be done with shorter exposure duration. Thus, for those biologic tissues, the phase contrast imaging can be better than the conventional absorption image. In the last decades, there have been already many methods to extract X-ray phase shift information [1]-[5]. However, the requirement of high brightness and the coherence of the X-ray source means that none of these methods can be applied into practice.

In 2006, a three-grating interferometry phase-sensitive imaging method [6] is proposed, which can lower the requirement of X-ray source. This method is shown in Fig.1. By grant G0, the normal laboratory incoherent X-ray source is split into several coherent line sources. These line sources can do phase contrast imaging separately. The role of G1 and G2 is to convert the phase shift information into the intensity of the X-ray.

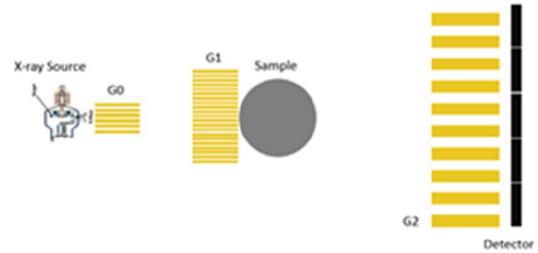

FIG. 1. Schematic diagram of the three-grating interferometry methods

The computed tomography (CT) is one of most effective way for 3D image reconstruction. NSRL (national synchrotron radiation laboratory) is developing a high phase contrast medical CT relied on three-grating interferometry. In order to obtain 200mm*mm of vision field, this prototype adopt 43 detector boards. Each detector board contain 384 channels coupled with a 20bit analog-to-digital converter (ADC). A trigger signal comes every 1.25ms to make 43 detectors start acquisition, which means the total bandwidth of 16512 channels is more than 250Mbps. In addition, as a spiral scanning device, the data of all detectors should be transferred from the rotator side to the stator side. Therefore, in this paper, we present a data transmission system for this prototype to solve the aforementioned problems.

## II. System Design

The data transmission system proposed in this paper contains two parts: three DCB and one DTB, as shown in FIG.2.

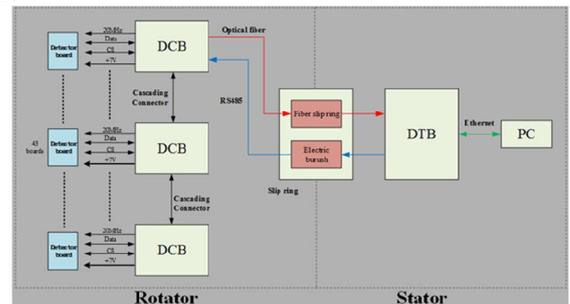

FIG. 2. Schematic Block diagram of the Data transmission System

Manuscript received Jun 15, 2018. This work was supported by National Key Scientific Instrument and Equipment Development Project under Grant No. ZDYZ2014-2.

Rongqi Sun, Lian Chen, Feng Li, and Ge Jin are with State Key Laboratory of Particle Detection and Electronics, University of Science and Technology of China, Hefei, Anhui 230026 China (phone:+86-0551-63606495,e-mail:srq@mail.ustc.edu.cn,chenlian@ustc.edu.cn,phonelee@ustc.edu.cn, goldjin@ustc.edu.cn)

Houbing Lu is with the College of Electronic Engineering,Natinal University of Defense and Technology, Hefei, Anhui 230037, China (e-mail: luhb@mail.ustc.edu.cn).

In this design, a slip ring including a fiber slip ring and an electrical brush is responsible for data uploading and command distributing from the rotator side to the stator side. On the rotator side, three DCBs act as the controller of 43 detectors. Each DCB is responsible for 15 detector boards. The design of one DTB is illustrated in FIG.3. The PCB of DCB is shown in the FIG.4.

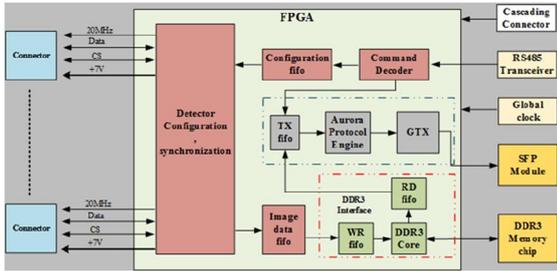

FIG. 3. Schematic Block diagram of the DCB board

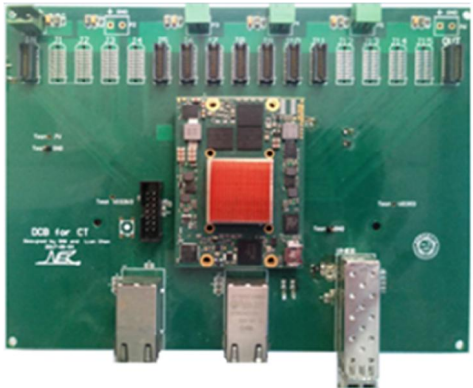

FIG.4. The PCB of the DCB.

Each DCB consists of 15 connectors, a DDR3 memory chip, a Kintex-7 FPGA, a RS485 transceiver a cascading connector and a SFP module. The core of the DCB board is based on a Kintex-7 FPGA, which is configured with a serial peripheral interface flash. In order to fulfill function of the communication between the rotator side and the stator side, a slip ring is used, which contains an electric brush and a fiber slip ring. A RS485 transceiver is responsible for the command distributing cross the slip ring, whose physical link is the electric brush. The communication between the DCB and one detector board is based on SPI protocol with 20MHz clock. When a negative pulse comes with the high chip select signal, the detector board starts to receive the configuration bits from the DCB. When the chip select signal is low, a handshaking signal makes the DCB receive image data from separate detector board. A check sum is appended to the image data packet. After checking is finished, the data from each detector are cached into a corresponding FIFO. When the full signals of all FIFO go high, these image data are read out to be store in the DDR3 memory chip sequentially. One NT5CC256M8FN DDR3 chip is adopted in the DTB board, which is driven by Xilinx MIG core. The speed is up to 1600Mbps. In order to implement the communication through an optical fiber, a so-called clock data recovery (CDR) technology has to been employed. The clock is inserted in the high speed bit streams and can be recovered by the receiver to latch the corresponding data. In our design, a gigabit transceiver is used to perform parallel-serial conversion (PCS) and high speed serial transmission. The Aurora 8B/10B protocol is a scalable, lightweight, link-layer protocol that can be used to move data point-to-point across one or more high-speed serial lanes, which is chosen in our design since there is no need of networking [7]. This Aurora 8B/10B protocol is configured in the simplex mode. As this is no back channel, a set of timers are used to drive the TX simplex initialization. Once a read command comes from the stator side, the image data from three DCBs are collected in the DDR3 of the first DCB by a cascading connector. Then all image data are read out and packed by Aurora 8B/10B protocol. After that, the encoded Aurora packets are serialized by the gigabit transceiver serializes. With the rising of commutation speed, the high frequency component attenuation of signals becomes a big problem. The GTX transceiver uses the pre-emphasis technology to compensate the high frequency attenuation. A SFP module does the electro-optical conversion and transmits these signals by an optical fiber. All control logic is integrated in the FPGA.

To simplify the design, the DCB and DTB share the same PCB with different FPGA logic. The schematic block diagram of the DTB is shown in FIG.5.

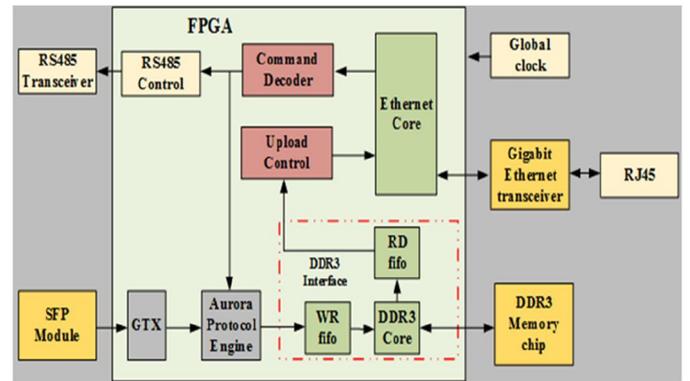

FIG. 5. Schematic Block diagram of the DTB board

The DTB board includes a SFP module, a kintex-7 FPGA, a DDR3 memory chip and Gigabit Ethernet transceiver working in the data link of the Ethernet. The FPGA is the controller of all peripheral device. The DTB is the TX side of command distributor. When PC send command and configuration bytes to DTB through a Gigabit Ethernet transceiver, FPGA is used to decode these instructions and distribute corresponding command to the rotator side by a RS485 transceiver. These serial image data received by the SFP module are fed into the GTX transceiver. After receiver equalization and word alignment by a comma code, these serial data are deserialized, then converted from 10bit to 8bit. The recovered parallel data are unpacked by Aurora 8B/10B engine. Generally, Aurora 8B/10B TX side re-initialize only when the 8B/10B decoding hard error event occurs in the RX side. Because there is no back channel available in this system between the rotator side and the stator side, the event-triggered re-initialization will not happen for most error. In this situation, a timer-triggered re-initialization mechanism is created in the

TX side, which will re-initializes and works again during a regular period. In order to judge whether any error occurred, in the user side, a CRC32 check is performed. If CRC32 check fails, a re-transmission command is given to the stator side by the RS485 transceiver. After CRC32 check, all image data packets are stored in the DDR3 memory chip and wait for uploading. The Ethernet interface is implemented in the MAC layer, which is sublayer of the data link layer [7]. In order to avoid the loss of the communication speed, there is no user layer protocol and network layer added in this design. The communication between PC and DTB is relied on the MAC layer directly and can achieve the interconnection by a regular switch. The WinPcap technology, a system framework of packet captures and network analyzers, is used for capturing row data and downloading command bytes [8].

### III. TEST RESULT

The code error rate of the optical fiber communication is evaluated, which is usually caused by the signal integrity and power instability. Inside the GTX, a pseudo-random binary sequence generator (PRBS) is placed between the CDR module and the word alignment module, which can be used to test the error rate. The IBERT tool provided by Xilinx is used in the test. The DTB and DTB are interconnected by an optical fiber, as shown in FIG.6. The speed is set to 1Gbps, which is match to practical application. The test result is illustrated in FIG.7. The code error rate is as low as 10E-12 and no bit error occurs during the test.

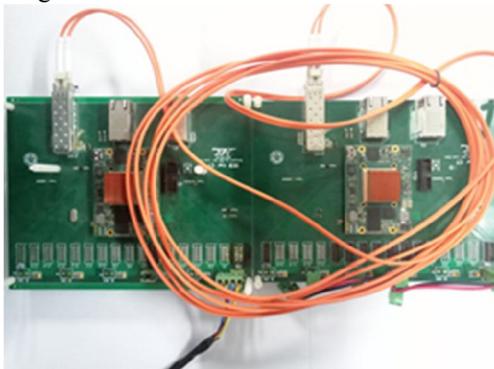

FIG. 6. The code error rate test

In order to do the Ethernet performance test, a test logic is integrated in the FPGA. When PC send a request to the DTB, the enable of linear data generator is open. The counter value consecutively is written into the TX FIFO at the speed of 1Gbps until the programmable threshold empty signal goes low. Then the data are uploaded into the PC by the Gigabit Ethernet transceiver. The reply for one request contains 256 packets. Each packet include 1024B. A long-time stability test for transmission rate is performed for 10 hours. The average Ethernet speed is calculated per 2 minutes. The test result is shown in FIG.8. The speed can reach up to 850Mbps.

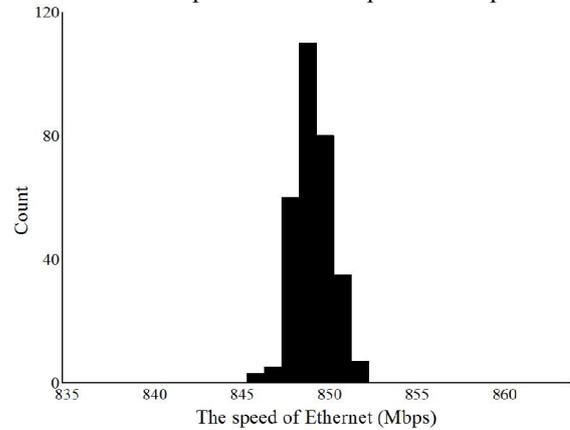

FIG.8. Long-term stability test of the gigabit Ethernet

### IV. CONCLUSION

This paper presents a data transmission system for a phase contrast X-ray human computed tomography prototype. This system is used to build test platform for theses detectors successfully. This data transmission structure can be a guidance for similar device.

FIG.7 The test result of IBERT